\documentclass{article}

\usepackage{arxiv}
\usepackage{amssymb}
\usepackage{algorithm}
\usepackage{algpseudocode}
\usepackage[utf8]{inputenc} 
\usepackage[T1]{fontenc}    
\usepackage{url}            
\usepackage{booktabs,footnote}       
\usepackage{indentfirst}
\usepackage{amsmath,amsfonts}       
\usepackage{microtype}      
\usepackage{lipsum}
\usepackage{graphicx}

\title{A Fully-Automated Framework Integrating Gaussian Process Regression and Bayesian Optimization to Design Pin-Fins}

\author{
  Susheel Dharmadhikari \\
  Department of Mechanical Engineering\\
  Pennsylvania State University\\
  State College, PA 16802 \\
  \texttt{sud85@psu.edu} \\
  \And
  Reid A. Berdanier \\
  Department of Mechanical Engineering\\
  Pennsylvania State University\\
  State College, PA 16802 \\
  \texttt{rberdanier@psu.edu} \\
  \And
  Karen A. Thole \\
  Department of Mechanical Engineering\\
  Pennsylvania State University\\
  State College, PA 16802 \\
  \texttt{kat18@psu.edu} \\
  \And
  Amrita Basak\footnote{Corresponding Author} \\
  Department of Mechanical Engineering\\
  Pennsylvania State University\\
  State College, PA 16802 \\
  \texttt{aub1526@psu.edu} \\
}

\begin{document}

    \maketitle
    \begin{abstract}
    Pin fins are imperative in the cooling of turbine blades. The designs of pin fins, therefore, have seen significant research in the past. With the developments in metal additive manufacturing, novel design approaches toward complex geometries are now feasible. To that end, this article presents a Bayesian optimization approach for designing inline pins that can achieve low pressure loss. The pin-fin shape is defined using featurized (parametrized) piecewise cubic splines in 2D. The complexity of the shape is dependent on the number of splines used for the analysis. From a method development perspective, the study is performed using three splines. Owing to this piece-wise modeling, a unique pin fin design is defined using five features. After specifying the design, a computational fluid dynamics-based model is developed that computes the pressure drop during the flow. Bayesian optimization is carried out on a Gaussian processes-based surrogate to obtain an optimal combination of pin-fin features to minimize the pressure drop. The results show that the optimization tends to approach an aerodynamic design leading to low pressure drop corroborating with the existing knowledge. Furthermore, multiple iterations of optimizations are conducted with varying degree of input data. The results reveal that a convergence to similar optimal design is achieved with a minimum of just twenty five initial design-of-experiments data points for the surrogate. Sensitivity analysis shows that the distance between the rows of the pin fins is the most dominant feature influencing the pressure drop. In summary, the newly developed automated framework demonstrates remarkable capabilities in designing pin fins with superior performance characteristics.
    \end{abstract}






\section{Introduction}

The hot-section components, e.g., turbine rotor blades of gas turbines operate at upwards of 1500 K creating a harsh environment. For this reason, innovative technologies are used to facilitate cooling of these components. A rotor blade typically has a complex cooling configuration. The blade can be divided into three primary sections, each of which has a dedicated coolant supply plenum. These sections include: (i) the leading edge region, (ii) the mid-chord region, and (iii) the trailing edge region. The pressure side of a turbine blade has a relatively higher temperature than the suction side and is, therefore, provided with additional cooling. This is accomplished by internal and external cooling that favors the pressure side of the blade, particularly in the trailing edge region \cite{town2018state}.

Internal cooling of the trailing edge has been investigated extensively. Heat transfer enhancement designs have been developed to fully utilize the cooling capacity of the coolant. Unique to this blade region is the pressure drop between the internal plenum and the external mainstream conditions. This region allows for greater heat transfer enhancement at the cost of increased frictional loss. Fully-bridged pin-fin arrays result in large pressure losses but also excel at heat transfer enhancement. As a result, pin-fins represent an ideal candidate for trailing edge cooling. As an additional benefit, the fully-bridged pin-fin design also increases the structural rigidity of the blade. For these reasons, a vast amount of literature supports the implementation of impingement arrays to enhance heat transfer and structurally support the trailing edge section of airfoils \cite{taslim2001experimental}.

One of the primary objectives of the pin-fin design is, therefore, to decrease the pressure drop while increasing the heat transfer. Both experimental and computational investigations have been carried out in the past to achieve this objective. Otto et al. \cite{otto2019vortical} performed a particle image velocimetry study to understand the developing flow characteristics of a staggered pin-fin array. Horseshoe type vortices and Karman instabilities were identified as the key contributors to turbulent mixing. Chyu et al. \cite{chyu2007comparison} investigated the dependence of pin-fin cross-sections on the thermal performance. Square, circular, and diamond-shaped pins were studied and circular pin-fins were found to have the best trade-off between pressure drop and heat transfer. 

With the advancements in metal additive manufacturing, the feasibility of making complex pin fin designs that are not limited to the traditional manufacturing constraints has increased multi-fold. This is evident from the recent research efforts towards the testing of such unique designs \cite{ferster2018effects}, and their corresponding parameters \cite{corbett2023impacts}. The introduction of this manufacturing technology has led to the initiation of research towards more innovative design strategies, particularly with the use of data-driven tools \cite{ghosh2020shape}. These strategies are built upon the parametrization of the pin fin designs followed by an optimization that leads to the desired pin fin shape.

Accordingly, there are several parameters pertaining to the pin shape that could be optimized to minimize pressure drop and maximize heat transfer. Existing literature has shown experimental results with unique geometries for pins such as a star or a dimpled sphere \cite{ferster2018effects}. However, experimental optimization of pin-fins is expensive and time-consuming. For this reason, computational investigations have played a crucial role in the development of novel designs and creative solutions. Eyi et al. \cite{eyi2019aerothermodynamic, eyi2019shape} used parameterized Bezier curves to define and, then, optimize the leading edge of a fin in a parametric form. Wileke et al. \cite{willeke2015adjoint} used adjoint optimization for a U-shaped channel to reduce the total pressure loss. On a similar note, Ghosh et al. and Dilgen et. al. implemented a topology optimization technique to explore manufacturing constraints \cite{ghosh2019topology, dilgen2018density}. Fabbri \cite{fabbri1997genetic} applied a genetic algorithm to optimize pin fin designs. Hamadneh et al. used particle swarm optimization (PSO) to evaluate several pin fin geometries for enhanced thermal performance \cite{hamadneh2013design}.

Recently, Ghosh et al. used Gaussian process (GP) surrogates with constrained Bayesian Optimization (BO) for optimizing the thermal performance of the pin-fin arrays \cite{ghosh2020shape}. Due to the black-box nature of CFD simulations for complex geometries, the use of GP and BO has shown promising results, particularly while working with limited data. However, due to the relatively nascent percolation of such techniques for pin fin optimization, more complex formulations in the design space are not yet fully explored. In addition to that, such studies have been hampered by the lack of automated simulations using established computational tools such as ANSYS Fluent, thereby, restricting the optimization to a few iterations. This paper addresses these existing shortcomings by applying a novel spline-based pin-fin definition that offers a potential to explore complex geometries. In addition to that, the simulations conducted in this analysis are completely automated leading to a relatively high number of design iterations. 

The optimization process in the present study is performed to minimize the pressure drop. The results reveal that the framework can learn the design principles with limited training data and converge to a desired solution with $<50$ iterations. A study on the data requirement of the algorithm is also conducted to quantify the need of initial data for the algorithm to perform adequately. Furthermore, a sensitivity analysis is presented to understand the impact of the features on the pressure drop. Although the analysis presented in the article is studied for minimizing pressure drop, it can be easily modified to address other desired objectives.

\section{Methodology}

\subsection{Featurization of Pin Fins}

\subsubsection{Features of Pin Fin Shape}

A general closed shape in 2D is a locus of points, $x = f(\theta)$, and $y = g(\theta)$, parameterized over $\theta$. The most common form of this parametrization is visualized with a circle of radius $r$, an outcome of $x = r\cos{\theta}$ and $y = r\sin{\theta}$. By employing more complex functional representations in the construction of $f(\theta)$ and $g(\theta)$, a range of variations in the shapes can be generated. Among several such strategies, this paper uses piecewise-cubic splines to generate parametric shapes in 2D. The complexity of these shapes, owing to their construction, further depends on the number of splines used in the process. 

\begin{figure}[!h]
    \centering
    \includegraphics[scale = 0.7]{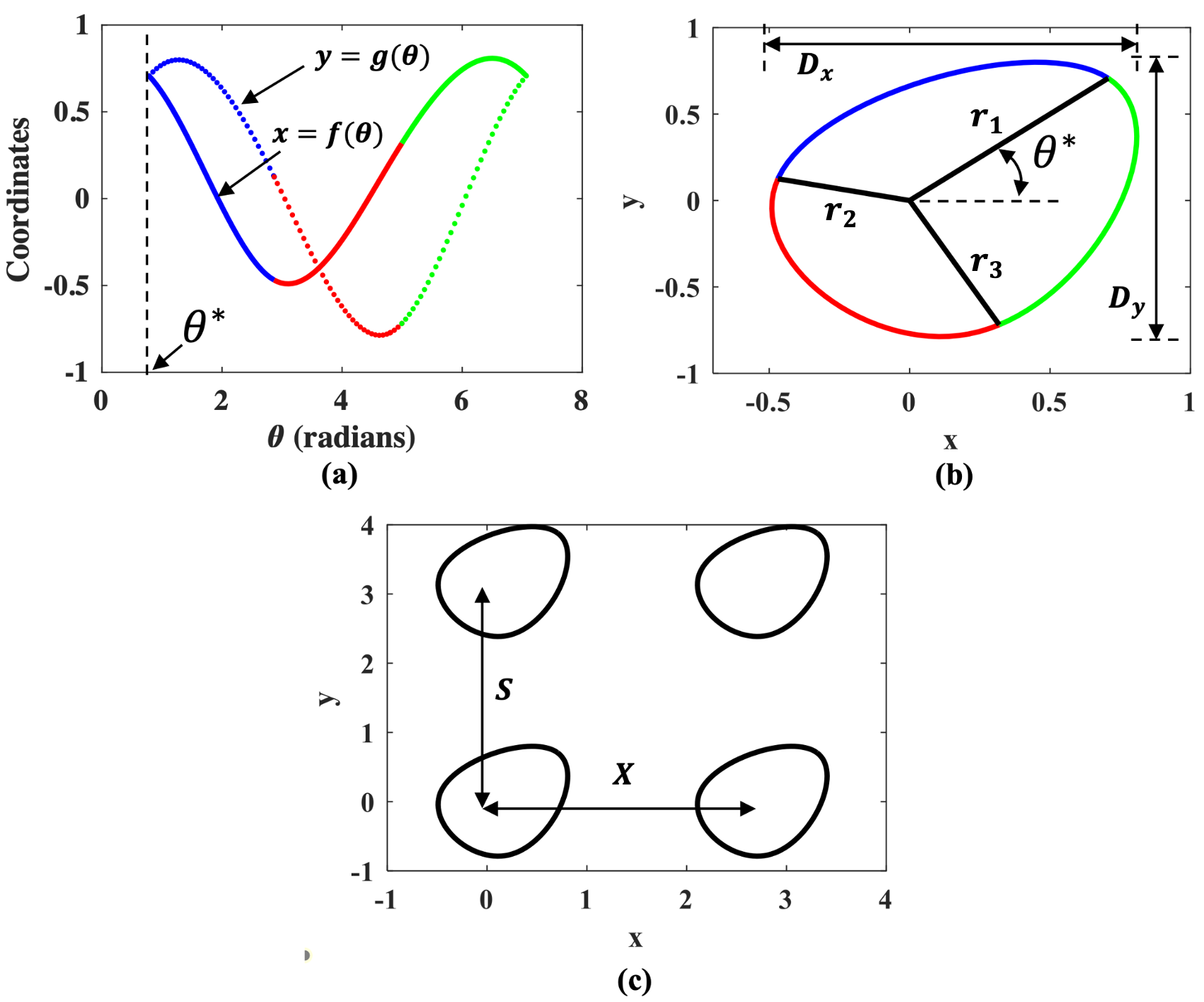}
    \caption{(a) Two curves ($x = f(\theta)$ and $y = g(\theta)$) constructed with three piece-wise cubic splines. (b) The pin-fin shape resulting from three splines with the defining parameters. (c) A pin fin array constructed with two rows and two columns.}
    \label{fig_params}
\end{figure}

With this in mind, the analysis in this paper is performed by using three splines. As an example, Fig. \ref{fig_params}(a) depicts the two curves ($x = f(\theta$) and $y = g(\theta)$), both constructed using three cubic splines. The resultant shape, indicating the contribution of the individual splines, is shown in Fig. \ref{fig_params}(b). The shape indicates three radial distances ($r_1$, $r_2$, and $r_3$) that provide the necessary coordinates for spline interpolation.

To be able to optimize this shape, features that impact the geometry need to be chosen. Three such features, viz. $r_2$, $r_3$, and $\theta^*$ are identified to be the defining elements for any shape generated using the aforementioned procedure. The feature, $r_1$, is maintained at a constant magnitude of 1 mm to ensure a reference dimension to prevent the optimization algorithm from choosing extreme (either too small or too large) geometries. The angle of $r_1$ from the X-axis is denoted by $\theta^*$, as shown in Fig. \ref{fig_params}(a) and (b). It controls the orientation of the pin fin. Fig. \ref{fig_params}(b) also depicts $D_x$ and $D_y$ which denote the projection length of the pin on X and Y axis, respectively.

\subsubsection{Features of Pin-Fin Arrays}

In addition to the three features of a single fin, the setup of the array is controlled with two additional features that account for the distance between the rows and columns of the fins. In literature, the distance between the rows is denoted by $S$, whereas, for columns, it is denoted by $X$ as shown in Fig. \ref{fig_params}(c). Accordingly, $S/D$ and $X/D$ are the two ratios that are commonly discussed in pin fin literature where $D$ is the projected length of the fin perpendicular to the flow. In this formulation, a minor variation of these ratios is used to avoid numerical inconsistencies in design. Instead of using $S/D$ and $X/D$, the authors define $S/D_y$ and $X/D_x$ as the two additional features based on their projection lengths from Fig. \ref{fig_params}(b). The inclusion of this modification avoids intersection of two fin shapes on the grid, which is observed to be happening in cases where the optimization converges toward fins with a high $D_x/D_y$ ratio.

Therefore, in total, the design ($\Omega$) of a pin-fin array can be uniquely defined by five features, viz. $r_2$, $r_3$, $\theta^*$, $S/D_y$, and $X/D_x$:
\begin{equation}
    \Omega = \left[ r_2, r_3, \theta^*, S/D_y, X/D_x \right]
\end{equation}

\begin{figure}[!h]
    \centering
    \includegraphics[scale = 0.7]{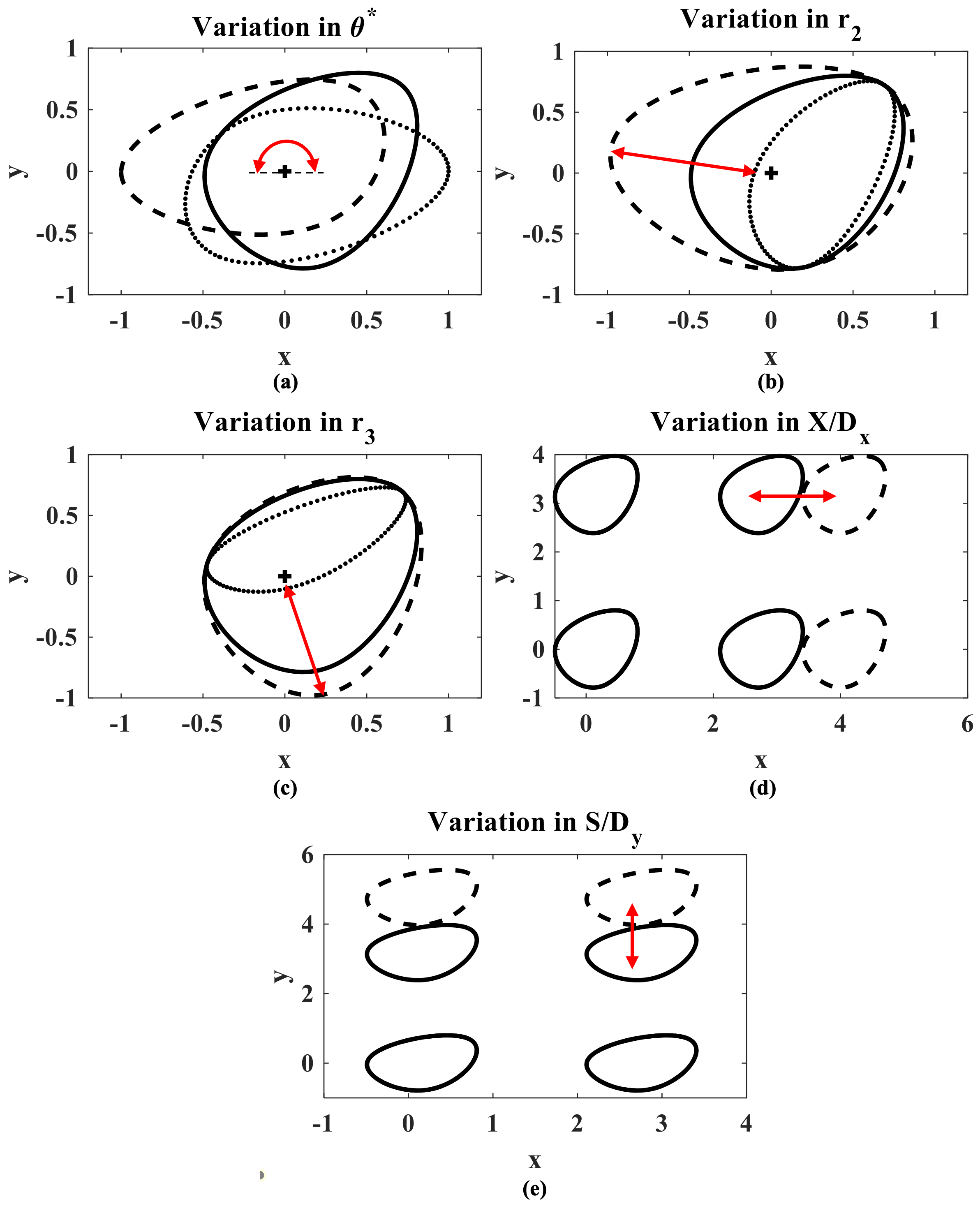}
    \caption{Feature space variation for (a) $\theta^*$, (b) $r_2$, (c) $r_3$, (d) $X/D_x$, and (e) $S/D_y$. The (red) arrows indicate the extent and the direction of variation of the parameter.}
    \label{fig_paramsvar}
\end{figure}

\subsubsection{Design Space}
To proceed with the optimization, some constraints on the range of variation for all the five features are required. Because a reference radial distance ($r_1$) is maintained at 1 mm, the other two radii $r_2$, and $r_3$ are varied from a lower bound of 0.1 mm to an upper bound of 1 mm. The lower bound is chosen to avoid numerical complications in spline computation which are observed for radii approaching zero. The impact of variation of these parameters on the fin shape is shown in Fig. \ref{fig_paramsvar}. 

As depicted in Fig. \ref{fig_paramsvar}, the orientation parameter ($\theta^*$) is varied from 0 to $\pi$ radians. Due to the nature of the flow, any pin fin design with $\theta^*$ has the same characteristics as $2\pi - \theta^*$. Therefore, the range $0$ to $\pi$ ensures all distinct orientations are taken into account. Every single variation between these three parameters $\theta^*$, $r_2$, and $r_3$ leads to a unique shape and thereby provides a multitude of design combinations. The variation in array parameters is relatively straightforward. Based on literature \cite{ferster2018effects}, the range of these parameters is chosen to vary from 2 to 3 and its impact on the array is shown in Fig. \ref{fig_paramsvar}(d) and (e). The design space is represented through a vector $\Omega^*$ as follows:

\begin{equation}
\begin{split}
    \Omega^* = & [ r_2 \in [0.1, 1] ,\\
             & r_3 \in [0.1, 1], \\
             & \theta^* \in [0, \pi], \\
             & S/D_y \in [2, 3], \\
             & X/D_x \in [2, 3] ]
\end{split}
\end{equation}

\subsection{Development of the Computational Fluid Dynamics (CFD) Model}

\subsubsection{Simulation Domain and Boundary Conditions}

\begin{figure}[!h]
    \centering
    \includegraphics[scale = 0.8]{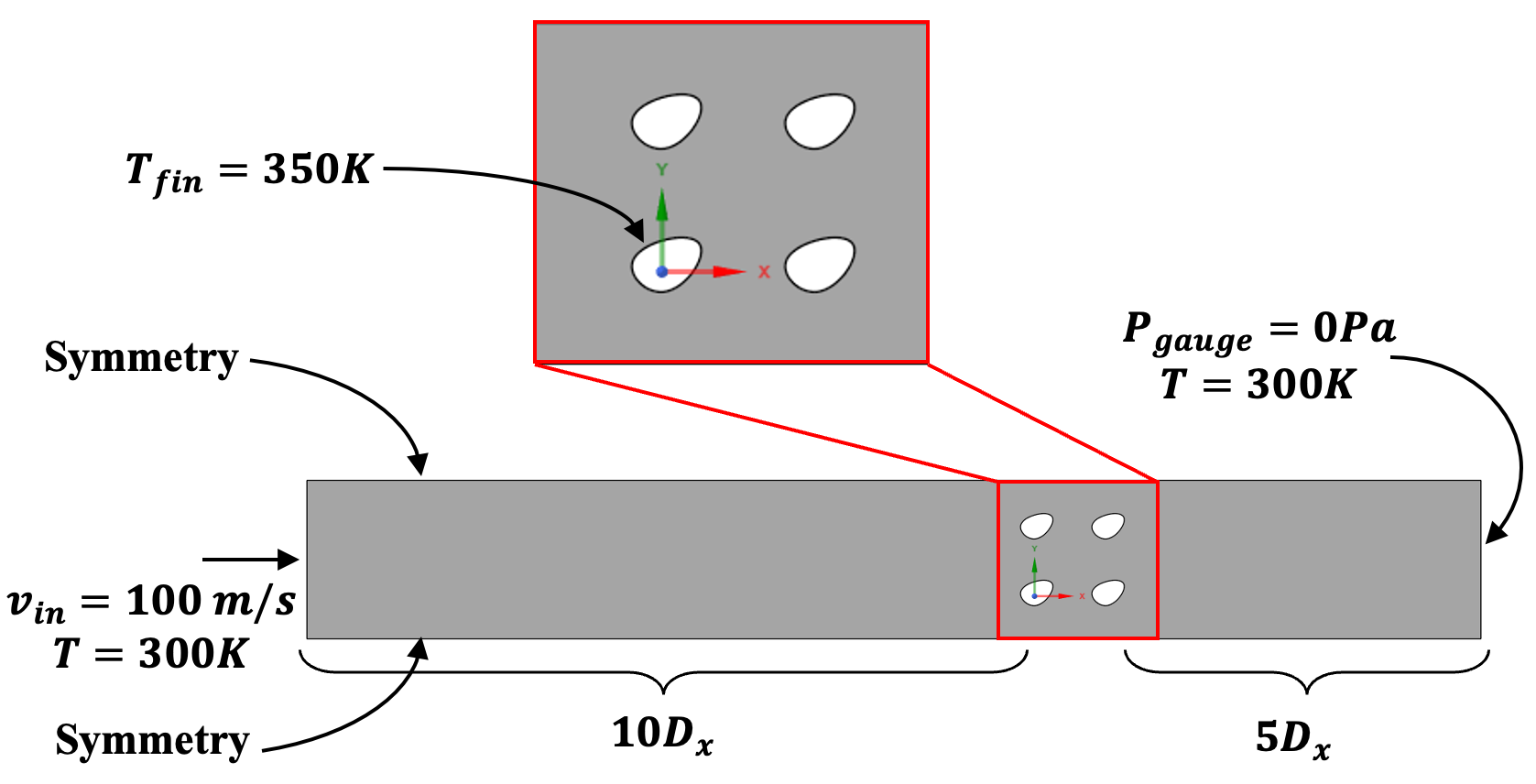}
    \caption{Simulation domain and boundary conditions implemented in the CFD model.}
    \label{figBC}
\end{figure}

The geometry of the 2D pin-fin array and the corresponding flow domain is shown in Fig. \ref{figBC}. The inlet passage of the domain is extended to $10D_x$ upstream to allow the flow to fully develop before interacting with the fins. An inlet velocity ($v_{in}$) of 100 m/s is maintained with a temperature of 300 K. The fins are modelled as a  stationary wall with a no slip condition and are maintained at a constant temperature ($T_{fin}$) of 350 K. The domain outlet is defined at atmospheric pressure ($P_{gauge} = 0$), with a temperature of 300 K, and the outlet passage is also extended to $5D_x$ to capture the flow behavior in the wake region. The top and bottom regions have a symmetry boundary condition. This ensures that any variation in the design space (for ex. due to $S/D_y$) would not have any impact on the simulations. The region of interest encapsulating the pins, shown in the inset in Fig. \ref{figBC}, will be shown in the results going further.

\subsubsection{Flow-Thermal Governing Equations}
The dynamics of the fluid flow is governed by the conservation of mass and the Cauchy momentum equation. In addition, the following assumptions about the physics of the fluid flow are made: 
\begin{enumerate}
    \item The fluid (i.e., air) is considered incompressible (with the constant density), thus the mass conservation translates into the volume conservation.
    \item The material properties of air are modeled with a linear constitutive behavior, i.e. a Newtonian fluid, where the internal shear stress is proportional to the shear rate.
    \item The fluid adheres to the surfaces of the fins and the walls; there is a no slip condition.
    \item The turbulence is specified using a 5\% turbulent intensity and a viscosity ratio of 10.
    \item In specifying the wall properties for the pin fin, a standard roughness model with a roughness constant of 0.5.
\end{enumerate}

Using the above assumptions, the time-averaged Reynolds-averaged Navier–Stokes (RANS) equations are used to describe the flow through pin fin arrays. The equations in Einstein notation for an incompressible Newtonian fluid and a stationary flow are written as:

\begin{equation}
    {\rho}\Bar{u}_j \frac{{\partial}\Bar{u}_i}{{\partial}x_j} = {\rho}\Bar{f}_i + \frac{\partial}{{\partial}x_j} \left[ -\Bar{p}\delta_{ij} + \mu \left( \frac{{\partial}\Bar{u}_i}{{\partial}x_j} + \frac{{\partial}\Bar{u}_j}{{\partial}x_i} \right) - {\rho}\overline{u_i^{\prime} u_j^{\prime}} \right]
\end{equation}

Here, $\rho$ is the density of the fluid, $\Bar{u}_j$ is the mean velocity in $j^{th}$ direction, and $\frac{{\partial}\Bar{u}_i}{{\partial}x_j}$ is the velocity gradient with respect to the $j^{th}$ direction. Therefore, ${\rho}\Bar{u}_j \frac{{\partial}\Bar{u}_i}{{\partial}x_j}$ is the change in the mean momentum of a fluid element owing to the convection in the mean flow. This change is balanced by the mean body force ${\rho}\Bar{f}_i$, the isotropic stress due to the mean pressure field $-\Bar{p}\delta_{ij}$, the viscous stresses $\mu\left( \frac{{\partial}\Bar{u}_i}{{\partial}x_j} + \frac{{\partial}\Bar{u}_j}{{\partial}x_i} \right)$, and apparent stress ${\rho}\overline{u_i^{\prime} u_j^{\prime}}$ owing to the fluctuating velocity field, generally referred to as the Reynolds stress. Here $f_i$ represents the external force, $\Bar{p}$ is the average pressure, $\Bar{u_i^{\prime}}$ denotes the mean of the fluctuating component of the velocity. For modelling turbulence, the SST $k-\omega$ formulation is used \cite{lynch2011computational}.

One of the future objectives of this research is to perform optimization studies that simultaneously minimize pressure drop and maximize heat transfer in a pin-fin array. Hence, in the present research the energy conservation equations are also solved. The transport of thermal energy is computed using the following equation:

\begin{equation}
    \frac{\partial (\rho E)}{\partial t} + \nabla \cdot [\textbf{V} (\rho E + p)] = \nabla \cdot \left[ k_{eff} \nabla T + \tau_{eff} \cdot \textbf{V} \right]
\end{equation}

Here, $E$ is the energy per unit mass, $\textbf{V}$ is the velocity vector, $p$ is the pressure, $k_{eff}$ is the effective thermal conductivity, and ${\tau}_{eff}$ is the effective shear stress.

\subsubsection{Model Implementation}
\begin{figure}[!h]
    \centering
    \includegraphics[scale = 0.6]{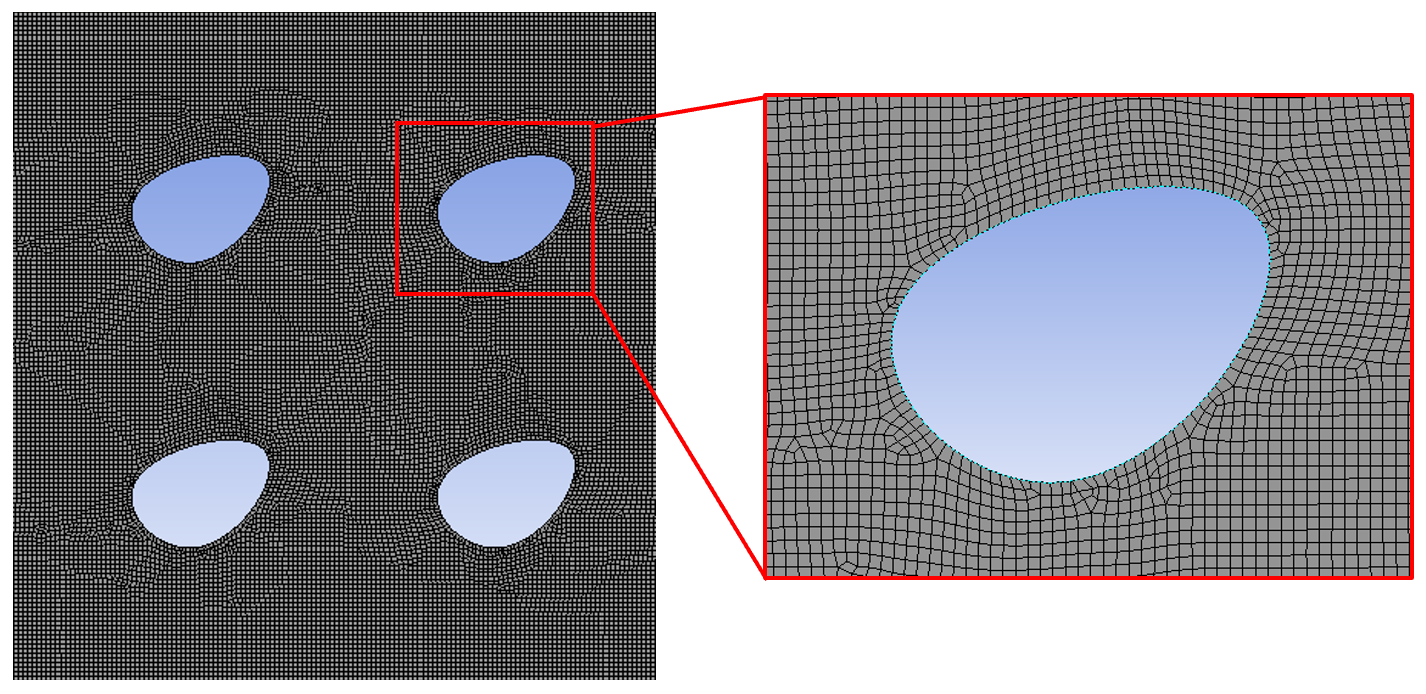}
    \caption{Mesh distribution near the fin boundaries.}
    \label{figmesh}
\end{figure}

The fluid flow and thermal evolution are simulated with the software ANSYS$^{\text{\textregistered}}$ Fluent 2020 R2, which is based on the finite-volume method. The governing equations are integrated over a finite set of quadrilateral control volumes that meshes the simulation domain. Details of the mesh are represented in Fig. \ref{figmesh}. Following a cell-centered discretization, the numerical solver computes discrete values of the continuous velocity and pressure fields at the center of the control volumes. The values at any other locations are interpolated from the discrete values, whenever required. The numerical scheme evaluates the advection and diffusion fluxes of momentum through all the faces of the control volumes. Then, the accumulated quantity of momentum inside each control volume is updated according to its net fluxes. As the fluid is incompressible, the pressure field is a result of the continuity constraint. A pressure equation is derived from the law of mass conservation.

The evolution of the system is solved incrementally with an automatic time stepping method using pseudo transient settings. The simulation is run for 200 iterations which is observed to be a sufficient duration for model convergence. In the simulations, the maximum element size is maintained at 0.05 mm. The convergence conditions are set at $1e-6$ for all residuals. The discretized momentum equations and pressure equations relative to the set of control volumes are solved with an implicit solver that ensured stability of the numerical scheme. At each incremental time step, the numerical algorithm computes the new values of the primary discrete variables that are the local velocities, the pressure and temperature. Secondary results, such as the streamline of the flow, the shear rate, the viscous stress, or fluxes, are computed from the primary variables. The pressure drop (${\Delta}P$), which represents the critical objective for optimization in this study, is calculated by simply recording the inlet pressure at the end of the simulation (since the outlet is maintained at $P_{gauge} = 0$ Pa).

\subsection{Optimization Framework}

\subsubsection{Gaussian Process-Based Surrogate}
The surrogate development strategy is based on a class of stochastic processes called Gaussian Processes (GPs) that assume any finite collection of random variables to follow a multivariate jointly Gaussian distribution. For a finite collection of \textit{n} designs, $\underline{\Omega}$, the corresponding function outputs, $\underline{\phi}$ are assumed to have a multivariate jointly Gaussian distribution,
\begin{equation} 
    \label{eq:GP_formulation}
    \underline{\phi} \sim \mathcal{N} (\underline{m}(\underline{\Omega})\;, \underline{K}(\underline{\Omega}, \underline{\Omega^{\prime}})) 
\end{equation}

Here, $\mathcal{N}$ implies a Gaussian distribution. The underlying GP is completely characterized by a mean function: $\underline{m}(\underline{\Omega}) = E[\underline{\phi}]$, and a covariance function: $\underline{K}(\underline{\Omega},\underline{\Omega}^{\prime}) = E[\underline{\phi} - \underline{m}(\underline{\Omega}))(\underline{\phi}^{\prime} - \underline{m}(\underline{\Omega}^{\prime}))]$~\cite{rasmussen2003gaussian}. Here, $E[\star]$ denotes the expectation of $\star$. $\underline{\Omega}^{\prime}$ and $\underline{\phi}^{\prime}$ denote a set of finite designs other than $\underline{\Omega}$ and the corresponding functional output of it, respectively.

In order to understand the application of surrogate modeling, consider the situation where $n$ designs denoted by the $\underline{\Omega}$ are being computationally evaluated to generate the outputs $\underline{\phi}$. Using this data, a surrogate model can be established with the multivariate Gaussian formulation. The surrogate model, can now be used to estimate the output of a new design $\Omega^{n+1}$ using the following formulation for a conditional distribution:

\begin{equation}
{{\phi}^{n+1}}|{\underline{\phi}},{{\phi}^{n+1}},\Omega \sim \mathcal{N}(m^{n+1},K^{n+1})
\end{equation}

Here,
\begin{equation}\label{eq:eq2}
\begin{split}
 m^{n+1}=\underline{K}(\Omega^{n+1},\underline{\Omega}) \underline{K}(\underline{\Omega},\underline{\Omega})^{-1}\underline{\phi}
 \end{split}
\end{equation}

\begin{equation}\label{eq:eq3}
\begin{split}
 {K}^{n+1} = \underline{K}({\Omega}^{n+1},{\Omega}^{n+1}) -
\underline{K}({\Omega}^{n+1},\underline{\Omega})\underline{K}(\underline{\Omega},\underline{\Omega})^{-1} \underline{K}(\underline{\Omega},{\Omega}^{n+1})  
\end{split}
\end{equation}

Here, $\underline{K}$ is the covariance matrix. Thus, the predicted posterior distribution of the outputs at every test data point is also a Gaussian distribution, characterized by the mean, $m^{n+1}$ and covariance, $K^{n+1}$. A detailed mathematical account of GPs can be found in \cite{rasmussen2003gaussian}.

\subsubsection{Bayesian Optimization}
The term \textit{{optimization}} is used to denote minimization of an objective function. A maximization problem can be posed similarly by taking the negative of the objective function, $\phi$. To minimize $\phi$ over its domain, the solver needs to find:
\begin{equation}
     \hat{{\Omega}} = \operatorname*{argmin}_{{\Omega}\in {\Omega^{*}}}\phi({\Omega})
\end{equation}

Here, `argmin' finds the argument that gives the minimum value of $\phi$. The functional form of $\phi$ is typically unknown and, hence, a gradient-free or \textit{black-box} optimization is often utilized. BO is one such \textit{black-box} optimization technique~\cite{frazier2018tutorial} that leverages the predictions through a surrogate for sequential \textit{active learning} to find the global optima of the objective function. The active learning strategies find a trade-off between \textit{exploration} and \textit{exploitation} in possibly noisy settings \cite{frazier2018tutorial}, which facilitates a balance between the global search and local optimization through \textit{acquisition functions}. One commonly used acquisition function in BO is Expected Improvement (\textit{EI}).

The objective function, $\phi$, expressed as a GP, yields a posterior predictive Gaussian distribution characterized by the mean $m(\Omega)$ and standard deviation $K(\Omega)$ for $\Omega\in {\Omega^{*}}$, where $\Omega^{*}$ is the search space of the optimization challenge. The optimization algorithm proceeds sequentially by sampling $ \hat{\Omega} = {argmax}_{\Omega}EI (\Omega)$ at every step of the iteration process to add on to the dataset, after which the GP surrogate is retrained with the new data set to predict the acquisition potential for the next iterative step. This process continues until an optimum is reached, or the computational budget is extinguished. Since the acquisition potential is predicted over the entire search space by the surrogate, BO can achieve fast predictions without a lot of function calls in the search space (i.e., without having to run the simulations to obtain the objectives at all the search locations). This process otherwise, might be computationally infeasible when the search space is high-dimensional and the simulations are expensive.

\subsection{Implementation of Iterative and Automated GP and BO}
\begin{figure}[!h]
    \centering
    \includegraphics[scale = 0.8]{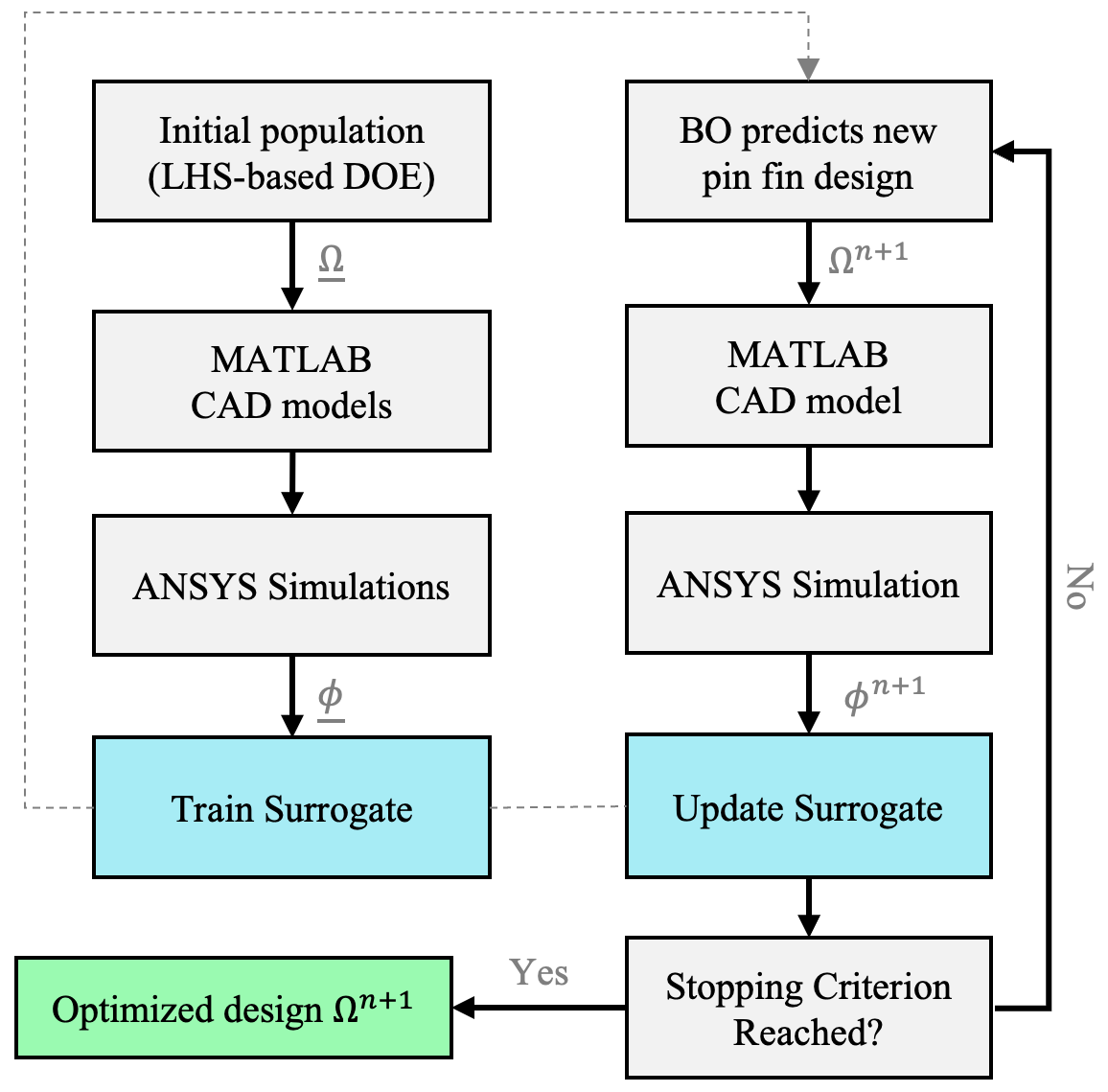}
    \caption{Flow chart of the optimization framework.}
    \label{fig_flowchart}
\end{figure}

The implementation of the proposed optimization framework is hinged on training and updating a surrogate model. Fig. \ref{fig_flowchart} depicts the workflow of the algorithm during the training and updating phase. In the training phase, a surrogate model is trained using an initial design population that is generated through a Latin hypercube sampling (LHS)-based design-of-experiment (DOE). For every design in the population a CAD model is generated in MATLAB, followed by an ANSYS simulation. Based on these outputs, a GP-based surrogate is trained. This surrogate forms the basis of the BO framework that again consists of three main computation aspects that need to be operated in sync iteratively. These aspects are (i) numerical pin fin shape generation and translation to a CAD geometry, (ii) CFD model setup and simulation, and (iii) Iterative BO using steps (i)-(ii). To achieve these steps, multiple softwares are operated through a master script in Python. The pin fin shapes are generated in MATLAB using the spline toolbox. The shape is then converted to an AutoCad file (.dxf) from MATLAB using the open source library \texttt{DXFLib}. The geometry is imported to ANSYS and the simulation is setup using pre-recorded journals. The output of the simulations is read from the simulations backup files by the python masterscript and is used to update the Bayesian optimizer until convergence or maximum specified iterations. The BO algorithm is employed using the \texttt{GPyOpt} toolbox. The entire framework is run without any manual intervention.

\section{Results and Discussion}

\subsection{Validation of the CFD Model}
\begin{figure}[!h]
    \centering
    \includegraphics[scale = 0.8]{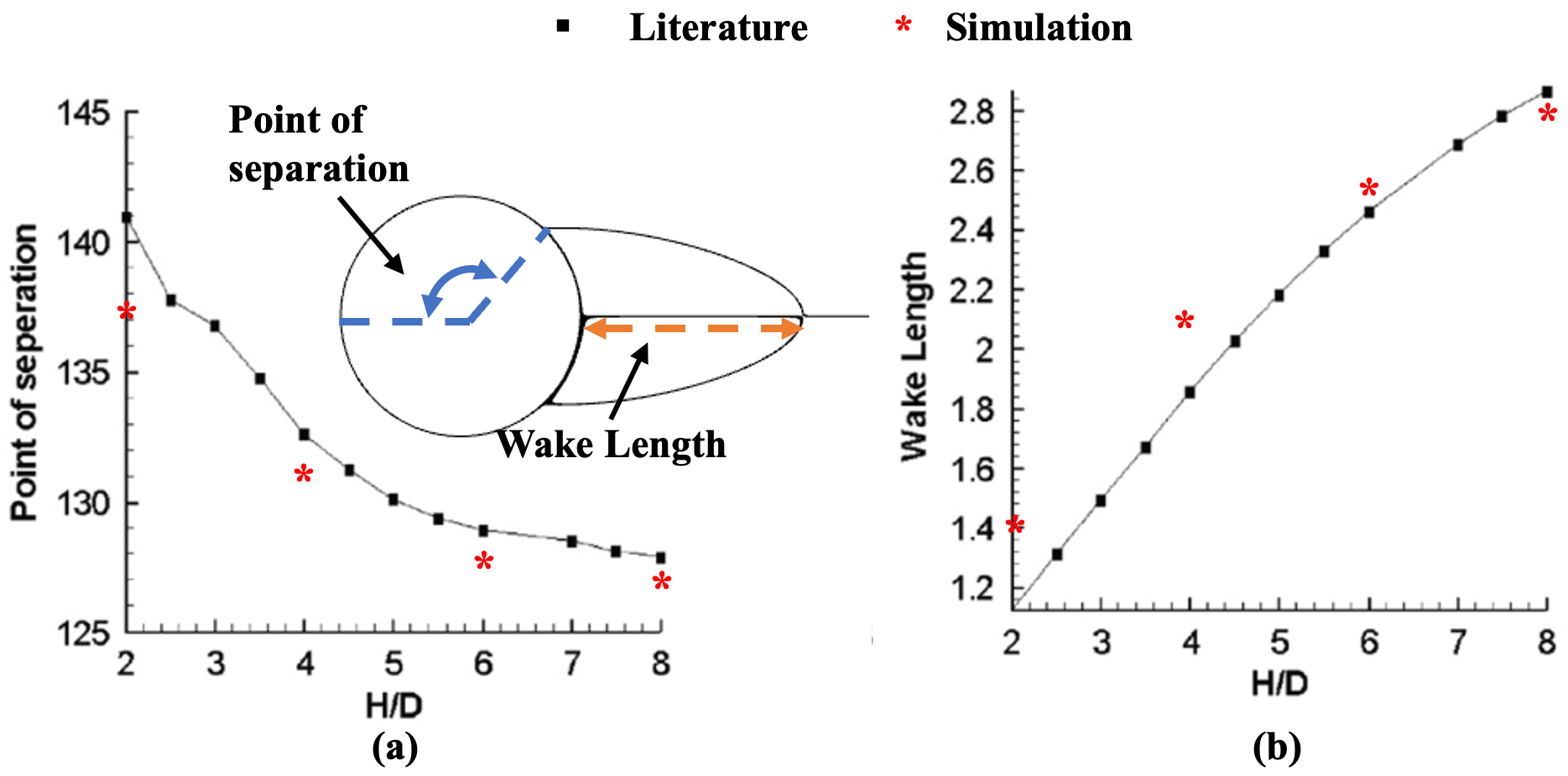}
    \caption{Comparison of the (a) point of separation and (b) wake length with published literature \cite{singha2010flow}}
    \label{fig_validation}
\end{figure}

Grid convergence is a necessary test in CFD simulations. In this study, the variation in the performance parameters e.g., ${\Delta}P$ is studied by altering the element size from 0.5 mm to 0.03 mm. This leads to a variation of 3,471 to 0.8 million nodes. All parameters, except the element size, are kept identical for all simulations. The variation in pressure drop is not significant (2\%) beyond 0.05 mm element size. Therefore, the element size of 0.05 mm is chosen for all analyses. To verify the setup of the Fluent module, a study is performed to compare the wake length and the point of separation to similar published research \cite{singha2010flow}. The comparison is shown in Fig. \ref{fig_validation} indicating that the Fluent module is adequately set.

\subsection{Performance of the Gaussian Process (GP) Surrogate}

\begin{figure}[!h]
    \centering
    \includegraphics[scale = 1.1]{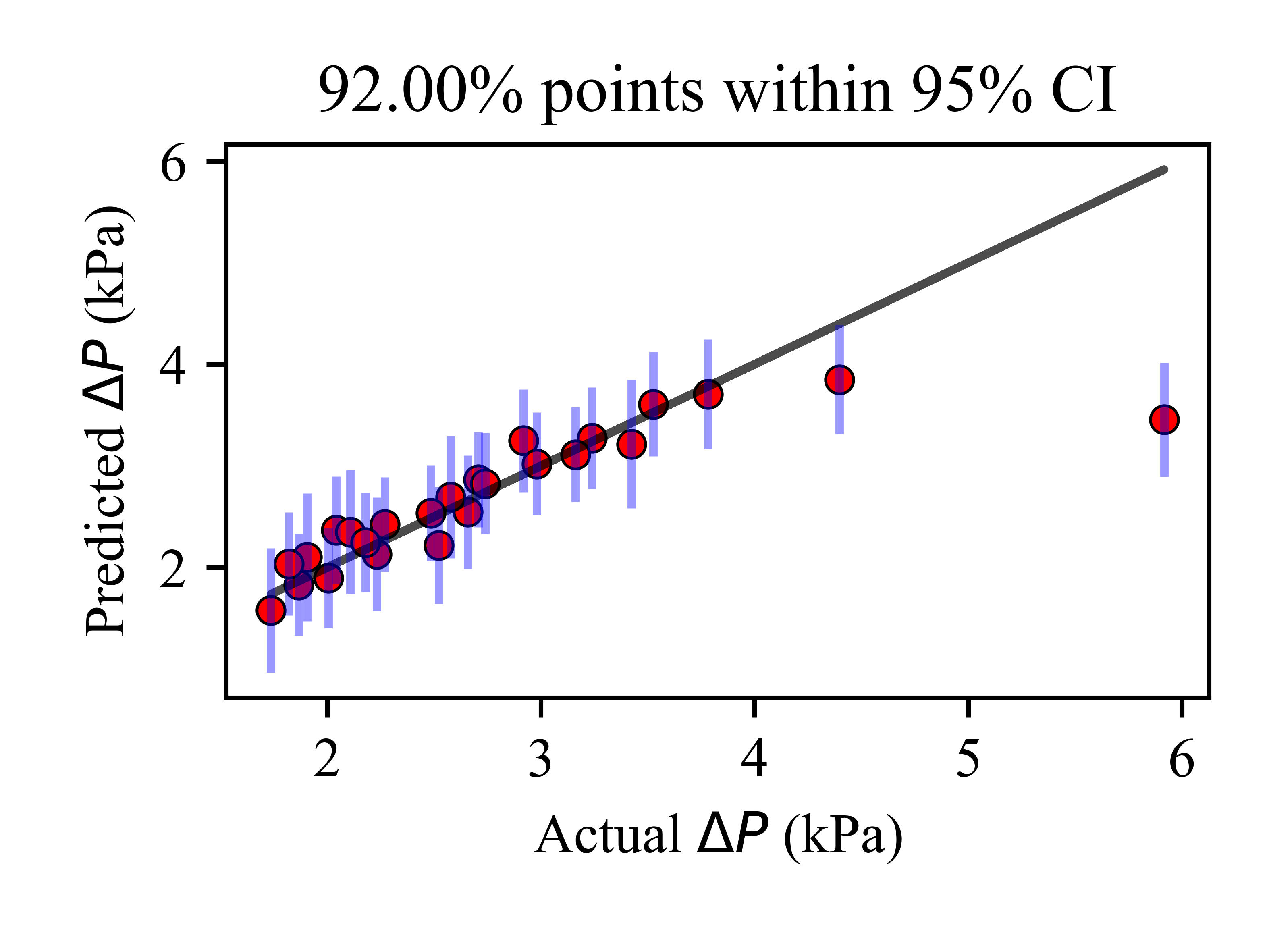}
    \caption{Comparison of ${\Delta}P$ between the prediction of the GP-based model against actual data.}
    \label{fig_gpperformance}
\end{figure}

Before moving to optimization, an LHS-based DOE is conducted with 100 designs to develop a surrogate. This surrogate forms the basis of the Bayesian optimization framework, and an efficient search of the optimal design depends on the construction of this model. Therefore, before diving into optimization, it is often advisable to check the accuracy of the surrogate using some regression metrics. In this study, to test the surrogate, the available data is randomly split using a 75\%-25\% ratio into a training and testing set. The training set is used to build the surrogate and the testing set is used to assess it. The result of the predictions against the actual data is shown in Fig. \ref{fig_gpperformance}. The model predicts 92\%  of the testing data within the 95\% confidence interval indicating that the surrogate is capable of emulating the actual physics. The Pearson correlation (R-squared), however, is low at 0.67. There is one conspicuous outlier in the data that shows a ${\Delta}P$ of 6 kPa. By removing that outlier, the model is capable of achieving an almost perfect accuracy. However, since the reasons for the outlier are enmeshed in the physics of the system, it is not removed during the optimization computations. It is also important for the GP to have some data for worst designs which helps in avoiding those design combinations later during optimization. The choice of 100 for the initial designs to build the surrogate is random. With optimization problems that are computationally expensive, 100 initial simulations already pose a challenge, and therefore, it is also essential to address these data requirements for the proposed algorithm to be successful. To address this issue, a discussion is followed in Section 3.4.

\subsection{Performance of Bayesian Optimization (BO)}

\begin{figure}[!h]
    \centering
    \includegraphics[scale = 0.9]{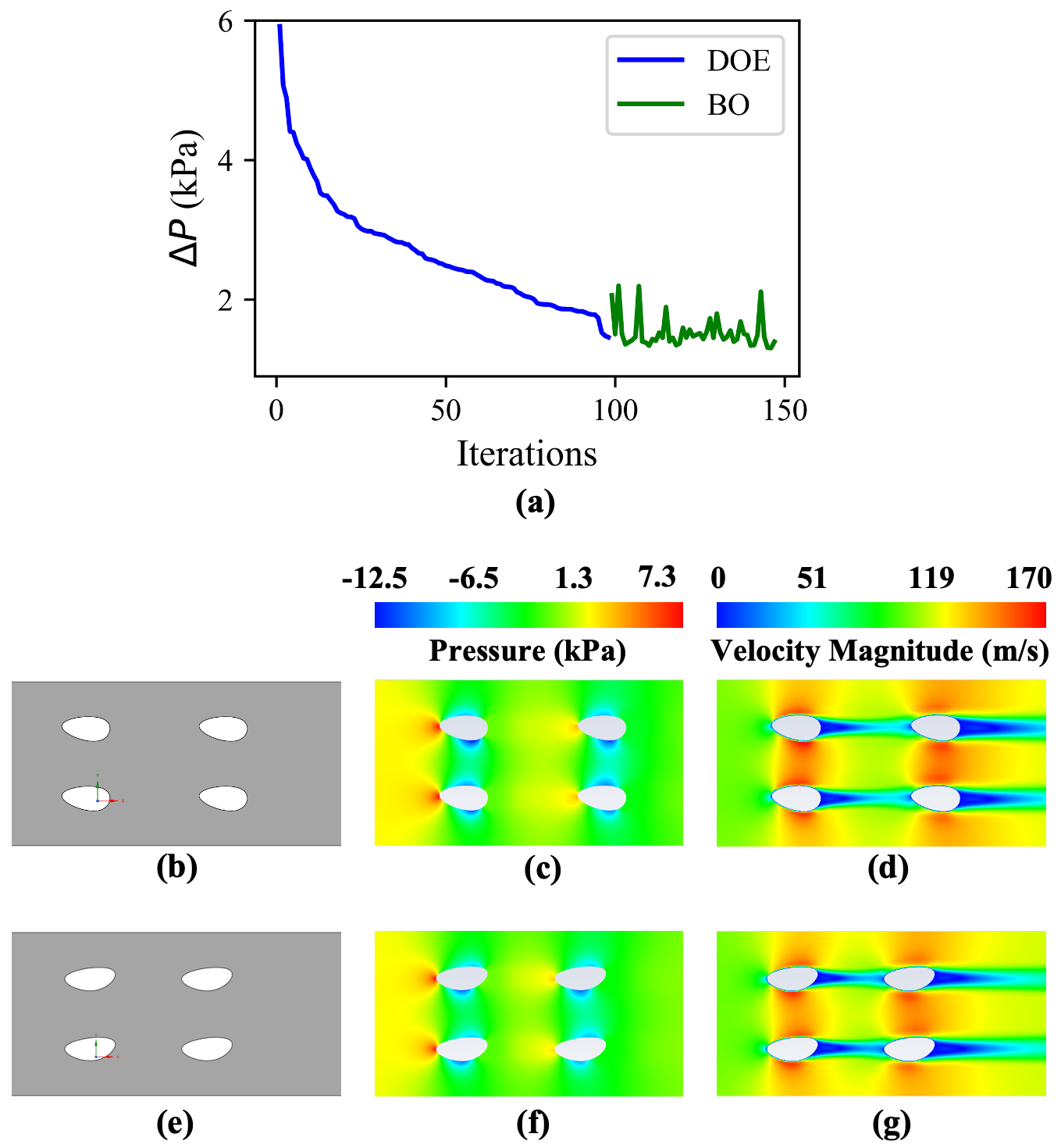}
    \caption{(a) Convergence of the BO algorithm. (b), (c), and (d) Best design, the corresponding pressure and velocity field, respectively, provided through the DOE. (d), (e), and (f) Optimized design, the corresponding pressures and velocity field, respectively.}
    \label{fig_boperformance}
\end{figure}

The convergence of the BO algorithm against the iterations and initial DOE information is shown in Fig. \ref{fig_boperformance}(a). The algorithm, that uses a surrogate built with 100 initial designs, finds an optimum within a few iterations, followed by an occasional (unsuccessful) exploitation of the design space indicated by the peaks in the convergence curve. The best design (Fig. \ref{fig_boperformance}(b)) with the corresponding pressure and velocity fields (Figs. \ref{fig_boperformance}(c) and (d)) provided by the initial DOE exhibits a ${\Delta}P$ of 1.46 kPa. The optimization algorithm is able to reduce it further to 1.3 kPa with a design (and pressure, velocity fields) shown in Figs. \ref{fig_boperformance}(d), (e), and (f). A comparison between the two designs in Figs. \ref{fig_boperformance}(b) and (e) reveal the impact of BO on making a more aerodynamic design resulting in an improved ${\Delta}P$.

\subsection{Evaluation of the Optimization Algorithm}
\begin{figure}[!h]
    \centering
    \includegraphics[scale = 0.8]{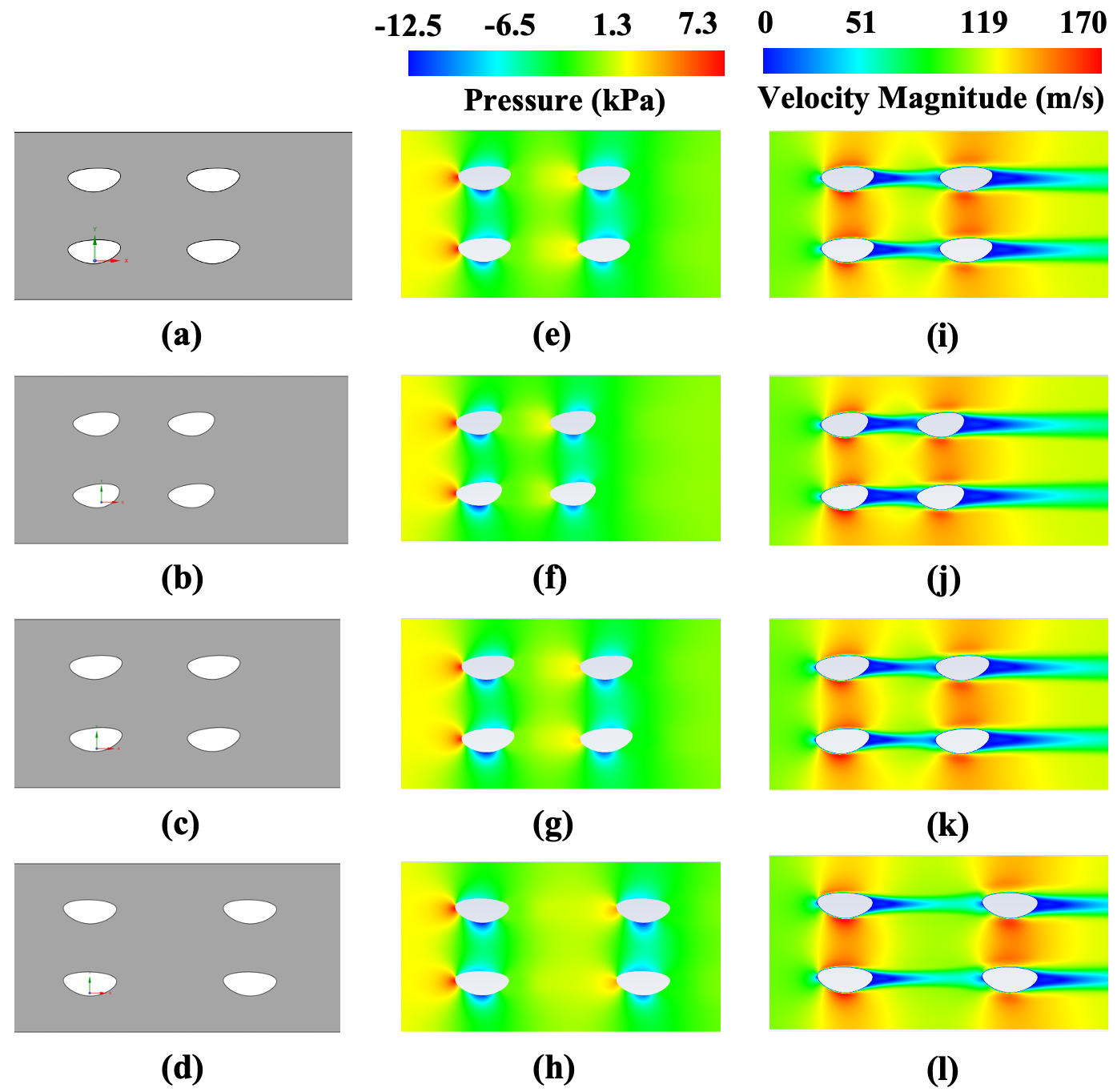}
    \caption{Optimized designs for cases with (a) 75, (b) 50, (c) 25, and (d) 0 initial designs prior to optimization, with their respective pressure fields in (e), (f), (g), and (h), and velocity fields in (i), (j), (k), and (l).}
    \label{fig_optimizeddesignslessdata}
\end{figure}

\begin{figure}[!h]
    \centering
    \includegraphics[scale = 0.85]{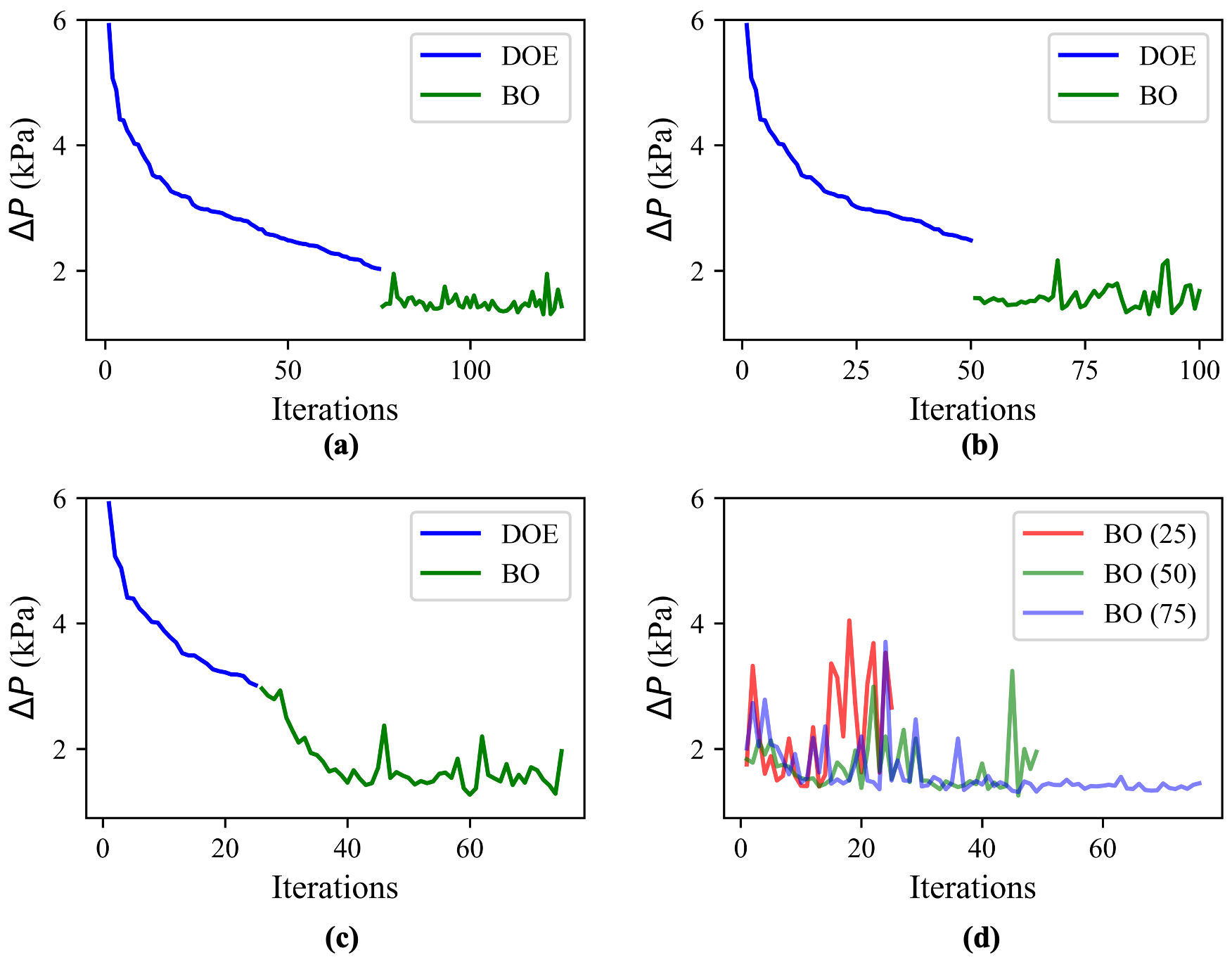}
    \caption{Convergence plot for BO with (a) 75, (b) 50, (c) 25, and (d) 0 initial designs.}
    \label{fig_convergencelessdata}
\end{figure}

The optimization result in the preceding section could reduce the pressure drop by 0.1 kPa using the information from 100 initial designs. In some practical scenarios, evaluating 100 simulations may not be feasible. Therefore, the capability of the BO algorithm to work with less information needs to be studied. In order to do that, the optimization is now carried out using four instances of less initial data by testing the functionality of the BO algorithm with 75, 50, 25, and 0 initial designs. The omission of the designs in each instance is such that the best designs are removed, thereby providing incrementally low information about the optimal solution to the BO algorithm. The best designs obtained through these simulations are shown in Fig. \ref{fig_optimizeddesignslessdata} and the convergence rates are shown in Fig. \ref{fig_convergencelessdata}. The features for all the optimized designs are tabulated in Table \ref{table_features}.

\begin{table}[!h]
\caption{Features of the optimized designs for different combinations of initial designs and BO steps}
\begin{center}
\label{table_features}
\begin{tabular}{c c c c c c c}
& & \\ 
\toprule
\# DOE Designs & $r_2$ & $r_3$ & ${\theta}^*$ & $S/D_y$ & $X/D_x$ & ${\Delta}P$ \\ [0.1cm]
  (BO Steps)       & (mm)  & (mm)  & (rad.)    &         &         & (kPa) \\ [0.1cm]
\toprule

100 (50)  & 0.1  & 0.74 & 2.8  & 3 & 2.3  & 1.3 \\ [0.1cm]
75 (50)   & 0.1  & 1.0  & 2.6  & 3 & 2.26 & 1.3 \\ [0.1cm]
50 (50)   & 0.17 & 0.75 & 2.8  & 3 & 2.04 & 1.3 \\ [0.1cm]
25 (50)   & 0.1  & 1.0  & 2.7  & 3 & 2.24 & 1.27 \\ [0.1cm]
0 (75)    & 1.0  & 0.1  & 0.49 & 3 & 3    & 1.25 \\ [0.1cm]
0 (50)    & 1.0  & 0.1  & 0.49 & 3 & 3    & 1.25 \\ [0.1cm]
0 (25)    & 0.62  & 1.0  & 2.86 & 3 & 2    & 1.4 \\ [0.1cm]
\hline
\hline
\end{tabular}
\end{center}
\end{table}

The optimized designs (Figs. \ref{fig_optimizeddesignslessdata}(a)-(d)) tend to approach a similar shape for all the instances indicating the presence of a global optima for this particular problem. The similarity in the performance of these designs can be further compared with the pressure and velocity fields in Figs. \ref{fig_optimizeddesignslessdata}(e)-(l). The subtle differences in them can be studied through the numerical values in Table \ref{table_features}. The designs, however, do not reveal the intricacies in which the algorithm approached the optima. That behavior is better exemplified by the convergence rates in Fig. \ref{fig_convergencelessdata}. With 75 initial designs (Fig. \ref{fig_convergencelessdata}(a)), the behavior of BO is almost similar to the previous case with all information (Fig. \ref{fig_boperformance}(a)). With 50 designs (Fig. \ref{fig_convergencelessdata}(b)), more understanding of BO can be inferred. By comparing Figs. \ref{fig_convergencelessdata}(a) and (b), one can notice that the first prediction of the BO algorithm is almost similar in both cases even after removing 25 best designs. This implies that the underlying GP learnt by BO with 75 and 50 designs is similar in its functional form.

This interpretation is further emphasized by comparing with Fig. \ref{fig_convergencelessdata}(c) where BO gradually moves towards an optimal design until 40 iterations, thereby indicating that the GP needed more that 25 designs to make a better informed decision. With the final case of 0 initial designs (Fig. \ref{fig_convergencelessdata}(d)), the convergence is not as steady as the previous cases. Since there is no initial data for this case, multiple instances with different limitations on maximum allowable iterations are conducted. The results for BO (25), BO (50), and BO (75) in Fig. \ref{fig_convergencelessdata}(d) exemplify the random nature of convergence for these simulations. Moreover, the intermittent peaks for ${\Delta}P$ that correspond to the exploitation phase in optimization have larger variance than the previous cases due to the unavailability of data. The predicted optima, however, is still close to the previous cases indicating the intelligent sampling procedure of BO. However, the predictions from such optimizations have a high probability of exploring local optima and are therefore unreliable. On an average, the BO algorithm is able to improve ${\Delta}P$ by more than 1 kPa as compared to the best design provided by the DOEs in all the cases.

\subsection{Sensitivity Analysis}
\begin{figure}[!h]
    \centering
    \includegraphics[scale = 0.7]{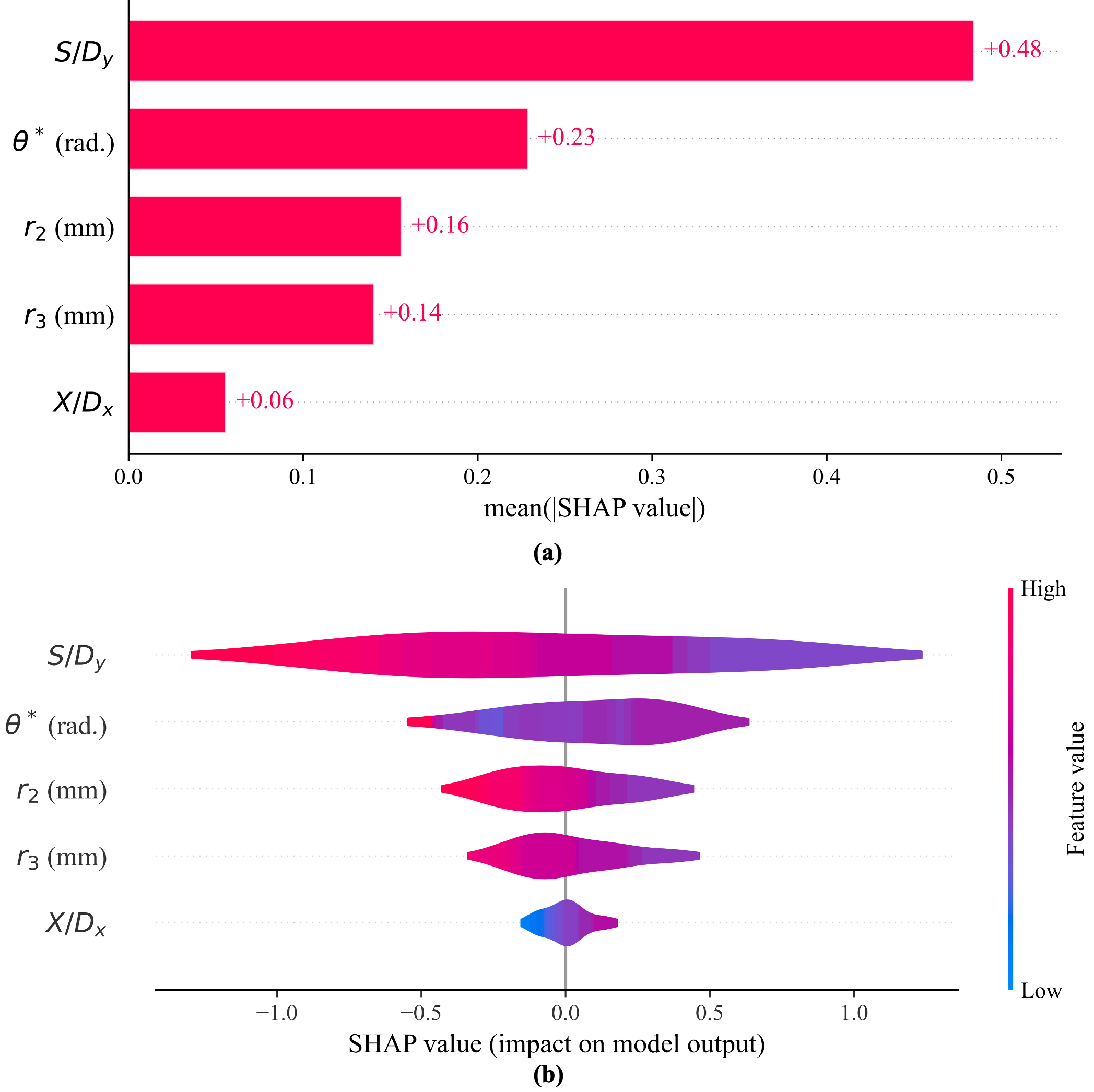}
    \caption{(a) A bar plot showing the relative absolute impact of the features. (b) A summary plot revealing the impact of the features on the output with respect to the changes in feature magnitudes.}
    \label{fig_shap}
\end{figure}

From a design and manufacturing point of view, it is essential to understand the relative impact of the features on the performance. Moreover, exploring the functional forms learnt by the GP can further help in understanding the system behavior. Hence, a global and local sensitivity analysis is now performed. The global analysis is essential to understand the impact of the features, whereas the functional forms from the GP can only be understood in a local context due to the multi-parametric nature of the problem. A SHAP (SHapley Additive exPlanations) analysis is performed to understand the global sensitivity of the features. SHAP is a method from coalitional game theory, developed to understand the individual impact of all the features in a prediction \cite{molnar2020interpretable}. Visually, the interpretation from this analysis can be presented in two forms, viz. (i) Using a bar chart as shown in Fig. \ref{fig_shap}(a), and (ii) Using a summary plot as shown in Fig. \ref{fig_shap}(b). Both the figures reveal important information about the feature behavior. 

The bar graph indicates the relative impact of the features on ${\Delta}P$. The X-axis of the graph shows the mean SHAP values that denote the average contribution of the features towards ${\Delta}P$. For example, for $S/D_y$, the mean SHAP value of 0.48 indicates that $S/D_y$ contributes to 0.48 kPa of the total ${\Delta}P$ predicted by the GP surrogate. Fig. \ref{fig_shap}(a) shows $S/D_y$ to be the most dominant factor influencing ${\Delta}P$, whereas $X/D_x$ has the least impact. Among $r_2$, $r_3$, and ${\theta}^*$ (the three features that create a shape), ${\theta}^*$ has the largest influence on ${\Delta}P$. Although this information is useful, it is impossible from the bars in Fig. \ref{fig_shap}(a) to interpret \textit{how} these features impact the outcome. For example, the bar chart does not tell whether increasing or decreasing $S/D_y$ is beneficial. This shortcoming is addressed through the summary plot in Fig. \ref{fig_shap}(b). The summary plot shows a scatter of color-coded violins across the X-axis for different features. The colors represent the relative magnitude of the features and the X-axis is the SHAP value. The length of the scatter indicates the relative influence. For example, for $S/D_y$, the scatter is the largest, indicating that it has the largest influence on the output. And the higher magnitudes of $S/D_y$ (red color) are towards the left end of the spectrum indicating that a higher $S/D_y$ would reduce ${\Delta}P$. This interpretation is also aligned with the optimized features (Table \ref{table_features}) where all designs have converged to the maximum possible $S/D_y$ to reduce ${\Delta}P$.

\begin{figure}[!h]
    \centering
    \includegraphics[scale = 0.8]{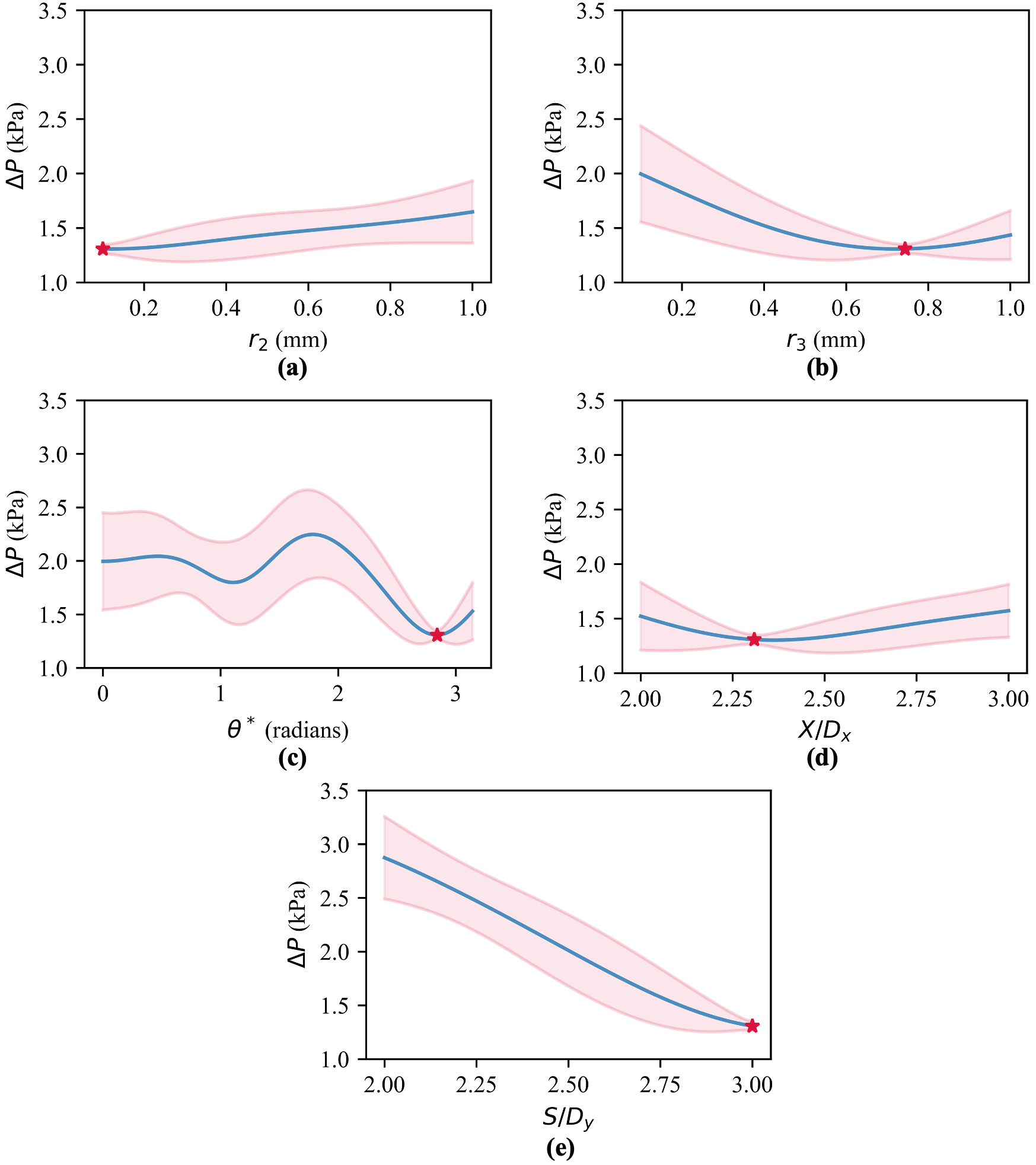}
    \caption{Local feature sensitivity on ${\Delta}P$ for (a) $r_2$, (b) $r_3$, (c) ${\theta}^*$, (d) $X/D_x$, and (e) $S/D_y$.}
    \label{fig_surrogates}
\end{figure}

To understand the local sensitivity, the behavior of the surrogate models is studied for the optimized design with 100 initial points. Fig. \ref{fig_surrogates} shows the variation of each feature with ${\Delta}P$ as learned by GP. To compute the variation for each feature, all other features are held constant at the optimized value indicated by BO. Therefore, the functional forms are heavily influenced by the constant feature values and the analysis is thereby termed as local. Even so, the variations are useful in understanding the impact on the optimized design. All the features, except ${\theta}^*$ show monotonic variation with ${\Delta}P$. The periodic variation in ${\theta}^*$ alludes to a symmetry that may be embedded in the CFD model. Among all the features, the relative total variation in ${\Delta}P$ indicates the impact of the feature on the outcome. As identified from the SHAP analysis, $S/D_y$ again causes the maximum variation in ${\Delta}P$ indicating its dominant impact. The star indicates the feature value in the optimized design. The sensitivity analysis therefore provides a comprehensive relationship between the objective and the features which ultimately aids in the design and manufacturing phases. The SHAP analysis provides a toolkit for varying features to satisfy the objective, i.e. setting a high value of $S/D_y$ in this case. Once an optimal design is selected, the local sensitivity analysis helps in identifying the features that need to be monitored (or controlled) more strictly than others depending on their impact on the outcome.

\section{Conclusion and Future Work}

The article presents a unique piece-wise cubic spline based framework for featurizing pin fins. An optimization problem for computing the pin fin arrays with minimum pressure drop is setup using a CFD model coupled with a surrogate-based Bayesian optimization approach. The optimized designs are observed to follow an aerodynamic shape leading to a reduction in the pressure drop. The capability of the BO framework is further tested with low initial information. The optimization is observed to efficiently find an optimum design with 25-50 initial data points. Furthermore, a sensitivity analysis is performed to reveal $S/D_y$ to be the most dominant feature to influence the pressure drop. Knowledge of the minimum number of designs needed for optimization coupled with the sensitivity analysis provide valuable information to design engineers.

The convergence to an aerodynamic shape with piece-wise cubic splines shows promise and will be explored further to test the capabilities of the method. With higher number of splines, more complex shapes emulating some of the tested prototypes \cite{ferster2018effects} can be generated. The mathematical setup of the pin fin designs also provides opportunities to include shape distortion that has been observed in additively manufactured specimens \cite{corbett2023impacts}. The studies with the modelling and impacts of such shape distortions will also be conducted for optimization. Geometrical constraints to compensate for these effects will make this approach more impactful and application-oriented. In addition to that, an extension of the method to three dimensions will also be pursued in the future. An imperative part of the current approach is the symmetry condition in the CFD model which in-theory implies infinite arrays of pins and an unbounded domain. To improve the predictions further, a bounded simulation emulating the actual testing environment will be conducted after finding the optimal pin fin shape. Moreover, the current method only tackles the pressure drop minimization problem. In the future, studies will also be conducted to perform a multi-objective optimization targeted towards enhancing heat transfer while reducing pressure drop. Experimental investigations will be performed to validate the efficacy of the newly developed framework and multi-fidelity modeling \cite{menon2022multi} will be pursued to intelligently blend experimental data with numerical data.

\section*{Credit Authorship}
Conceptualization, A.B., K.A.T., R.A.B., and S.D.; methodology, S.D.; software, S.D.; validation, S.D.; formal analysis, S.D.; investigation, S.D., A.B., K.A.T., and R.A.B.; resources, A.B.; data curation, S.D.; writing---original draft preparation, S.D., and A.B.; writing---review and editing, A.B., K.A.T., R.A.B., and S.D.; visualization, S.D.; supervision, A.B., K.A.T., and R.A.B.; project administration, K.A.T., and A.B.; funding acquisition, K.A.T., A.B., and R.A.B. All authors have read and agreed to the published version of the manuscript.

\section*{Acknowledgement}
The authors would like to thank Ritam Pal and Nandana Menon, PhD Students, Mechanical Engineering, Penn State for their help with CFD modelling and Bayesian Optimization, respectively, and Evan Mihalko, PhD Student, Mechanical Engineering, Penn State for proof-reading the manuscript.

\section*{Funding Information}
The research is funded by the NASA University Leadership Initiative program through grant number 80NSSC21M0068. Any opinions, findings, and conclusions in this paper are those of the authors and do not necessarily reflect the views of the supporting institution.

\section*{Data Availability Statement}
The data are available from the communicating author on reasonable request.

\section*{Conflicts of Interest}
The authors declare no conflict of interest.

 \bibliographystyle{unsrt} 
 \bibliography{V2}





\end{document}